\documentclass[12pt]{article}
\usepackage{amssymb,amsmath,epsfig}
\usepackage{cite}
\allowdisplaybreaks

\begin{document}
\title{\bf Stable Charged Radiating Systems Associated with Tilted Observers}

\author{Z. Yousaf \thanks{zeeshan.math@pu.edu.pk}\\
Department of Mathematics, University of the Punjab,\\
Quaid-i-Azam Campus, Lahore-54590, Pakistan.}

\date{}

\maketitle
\begin{abstract}
This paper is aimed to study the influence of electromagnetic field
and tilted congruences on the dynamical features of self-gravitating
system. We shall explore the stability of homogeneous energy density
in the background of Maxwell-Palatini $f(R)$ gravity. In this
respect, we have considered an irrotational non-static planar
geometry which is assumed to have two different types of gravitating
sources. The role of tilted congruences and the geodesic motion of
an evolving system is studied through the divergence of the entropy
vector field. The condition for the emergence of Minskoskian cavity
is also explored. In order to connect tilted and non-tilted
reference frames with inflationary and inverse Ricci scalar
corrections in the charged medium, few well-consistent relations are
presented. It is concluded that effective electric charge is trying
to increase the stability of regular energy density of the planar
system.
\end{abstract}
{\bf Keywords:} Gravitation; Self-gravitating systems; Relativistic
dissipative fluids.\\
{\bf PACS:} 04.20.Cv; 04.40.Nr; 04.50.Kd.

\section{Introduction}

General Relativity (GR) was the most influential gravitational
theory of the last era, widely understood as a theory explaining
geometrical attributes of the space and time on macroscopic scales.
The implication of GR provides Friedmann equations for a regular and
perfect fluid configurations that could accurately explain the
astrophysical transition of radiation and then matter dominated
cosmic epochs. Indeed, the current advancement of observational
cosmology accompanied by the highly precise experimentation, like
supernovae observations
\cite{pietrobon2006integrated,giannantonio2006high,riess2007new},
has revealed that our cosmos is in a state of accelerated expansion.
This phenomenon can not be explained by GR provided with the
conventional gravitating source.

In addition, GR does not take into consideration the cosmological
period known as inflation \cite{guth1981inflationary}, which was thought to have arisen before the radiation
stage and which could mitigate some of the challenges of standard
cosmology such as the flatness and horizon issues \cite{peebles1993principles}. Furthermore, GR with the
conventional baryonic matter could not address the observed fluid
density estimated by fitting the standard cosmic model with the
observations of WMAP7 \cite{komatsu2009five}.

Moreover, in order to study inflationary cosmic era in an
Einstein-$\Lambda$ gravity, one needs to add inflaton (slow roll
scalar field) by hand, thereby indicating that the cosmological
constant $\Lambda$ can not accommodate inflationary period of our
universe. Indeed, other interpretations for the aforementioned
acceleration can be given through theories which generalize GR by
adopting action function, dissimilar to the action of
Einstein-Hilbert. Nojiri and Odintsov \cite{nojiri2007introduction} addressed the
relevance and the need for these theories in depth. The $f(R)$ ($R$
is the Ricci scalar), $f(T)$ ($T$ is
the trace of energy momentum tensor), $f(R,\Box R, T)$ ($\Box$ is the de
Alembert's operator and $T$ is the trace of energy momentum tensor) etc., are among the appealing models of modified
theories (for further reviews on such models, see, for
instance,~\cite{NOJIRI2006144,Bamba2012,capozziello2011extended,bamba2013modified,
yousaf2016causes,yousaf2016influence,nojiri2007introduction,nojiri2008dark,nojiri2002friedmann}).
Numerous aspects of these theories
have also been extensively studied in gravitation and astrophysics
\cite{yousaf2019role}.

There are three different versions of $f(R)$ gravity. In this paper,
we are considering Palatini $f(R)$ gravity. In this theory while calculations, the
metric and connections are considered as independent quantities. The
Palatini $f(R)$ approach may lead to explaining several unusual
phenomenological GR modifications the study of supermassive compact
objects, structure formation and evolution of the universe
\cite{meng2004cosmological,allemandi2004accelerated,allemandi2007constraining,
sharif2015stability,olmo2011palatini}. This scheme of gravity provides
singularity free second order field equations instead of fourth
order, thus making them comparatively easy to solve and handle
mathematically \cite{sotiriou2006f}. Many scholars have addressed
the comprehensive analysis on the feasibility of this theory,
including Olmo and his collaborators
\cite{PhysRevD.72.083505,doi:10.1142/S0218271811018925,PhysRevD.86.044014,PhysRevD.86.104039,PhysRevD.86.127504}.
Recently, Olmo \emph{et al.} \cite{olmo2020stellar} described the challenges
as well as the achievements of modified gravity theories in
explaining the structure and evolution of stellar objects,
while junction condition for the matching of interior and exterior
metrics are found in \cite{olmo2020junction}

Ilyas \emph{et al.} \cite{Ilyas2017} studied the stellar evolution
by taking into account isotropic relativistic spheres and calculated
some feasible and stable models in modified gravity. Recently,
Moraes \emph{et al.} \cite{moraes2017analytical} investigated the
dynamics of compact objects after exploring modified versions of
equations of motion. Bhatti \emph{et al.} \cite{bhatti2020dynamical} after
applying numerical techniques calculated stable epochs of few
strange stars in modified gravity. Recently, few researchers have
calculated.  The astrophysical properties of various stellar bodies,
like wormholes \cite{sahoo2017wormholes, bambi2016wormholes},
gravastars \cite{yousaf2020construction, yousaf2020gravastars} and cosmic models \cite{yadav2020existence,malik2020study,yousaf2017stellar}
are also evaluated by many researchers in
modified gravity.

Research to investigate the explanation behind the phenomenon of
inhomogeneous energy density (IED) has motivated several theorists
not only in GR, but also in modified theories. To explore the
factors evolved throughout in the development of IED over the
stellar structures, Hawking and Israel\cite{hawking1979general}
found a physical relationship between the tidal forces and the
matter parameters. Herrera et al. \cite{herrera1997role} studied the
construction of naked singularity using IED and locally anisotropic
pressure for the spherically symmetric fluid configurations. Herrera
et al. \cite{herrera2004spherically} studied radiating spherical
compact objects in GR and predicted a remarkable connection between
tidal forces, IEDs, and anisotropic pressure. Furthermore,
Raychaudhuri evolution equation for the irrotational relativistic
spheres was found by Herrera et al. \cite{herrera2011role} via
well-known structural scalars. These scalars can be determined from
the orthogonal breaking down of the Riemann curvature tensor. The
role of structural variables on the dynamical instability and
irregularity factors of relativistic geometric populations with
matter distribution have been investigated by \cite{yousaf2017evolution}. They
concluded that the factors which control the stability and
inhomogeneities of the corresponding systems are energy density,
modified curvature terms and pressure.

Di Prisco \emph{et al.} discussed the solutions having thin shell
and also some solutions which satisfy the Darmois conditions on the
boundary $\Sigma$ \cite{di2011expansion}. Yousaf et al.
\cite{bhatti2017evolution,yousaf2018structure,yousaf2019role,RYousaf2019}
taken into consideration an/isotropic structure having the influence
of heat dissipation and analyzed its basic properties through
structure scalars. They formulated the field and dynamical equations
in terms of such scalars and described the importance of these
scalars in the modeling stellar bodies. Recently, Herrera
\cite{herrera2011tilted,herrera2017gibbs, herrera2020landauer} and
Yousaf \emph{et al.} \cite{PhysRevD.95.024024, yousaf2019tilted,
yousaf2019non} described the significance of tilted and non-tilted
observers in order to explain the some distinct physical properties
of the same matter configuration. Sussman and Jaime
\cite{sussman2017lemaitre} studied the class of inhomogeneous model
of non-interacting particles through LTB geometry in $f(R)$ gravity.
Yousaf \cite{doi:10.1142/S0217732319503334} calculated some
constraints describing inflationary and late time acceleratory
universe with the help of planar dissipative models with tilted and
non-tilted congruences.

This paper aims to describe the role of tilted congruences and
Maxwell-$f(R)$ gravity on the dynamical properties of planar
relativistic systems. We shall also present some relationship
connecting the matter variables of tilted and non-tilted observers
in the presence of electromagnetic field. The coming section is
devoted to present two different set of structural variables
corresponding to comoving and non-comoving frame of references.
Section \textbf{3} describes the Bianchi identities and few
well-known kinematical variables in the presence of Maxwell-$f(R)$
gravity. Furthermore, the factor involved in the emergence of IED is
also explored in the same section. The results are summarized in the
last section.

\section{Palatini $f(R)$ Formalism}

The action function for the derivation of field equation in $f(R)$
gravity can be given as follows
\begin{equation}\label{1}
S_{f(R)}=\frac{1}{2\kappa}\int d^4x\sqrt{-g}f(\hat{R})+S_M,
\end{equation}
where $f(\hat{R}),~S_M$ and $\kappa$ represent the generic function
of the Ricci scalar, action for matter and coupling constant,
respectively. Here $\hat{R} = g^{\gamma\delta}R_{\gamma\delta}$,
comes from the contraction of the $R_{\gamma\delta}$ and metric
tensors $g^{\gamma\delta}$ associated with the connection symbol. One should keep in mind that the Ricci scalar
as a function of the connection symbols (please see for details
\cite{PhysRevD.72.083505,doi:10.1142/S0218271811018925,PhysRevD.86.044014,PhysRevD.86.104039,PhysRevD.86.127504}).
The variations of the above action with respect to $g_{\gamma\delta}$ and
$\Gamma^\mu_{\gamma\delta}$, respectively provides
\begin{eqnarray}\label{2}
&&f_R(\hat{R}){\hat{R}}_{\gamma\delta}-[g_{\gamma\delta}f(\hat{R})]/2
={\kappa}T_{\gamma\delta},\\\label{3} &&
\hat{\nabla}_\mu(g^{\gamma\delta}\sqrt{-g}f_R(\hat{R}))=0,
\end{eqnarray}
where $f_R=\frac{df(R)}{d\hat{R}}$ and $T_{\gamma\delta}$ is an
energy-momentum tensor that is not dependent on connections. It
expression can be casted as
\begin{align}\label{2a}
T_{\alpha\delta}=-2(-g)^{-1/2}\frac{\delta S_M}{\delta g^{\alpha\delta}},
\end{align}
One can get the following equation after taking the trace of Eq.\eqref{2} as
 \begin{equation}\label{5}
\hat{R}f_R(\hat{R})-2f(\hat{R})={\kappa}T.
\end{equation}
thereby describing $R$ as dependent on $T$ and $T\equiv
g^{\gamma\delta}T_{\gamma\delta}$. The consideration of vacuum case,
i.e, $T=0$ in the above equation makes $R$ as a constant entity,
thereupon demonstrating it as function of $f(\hat{R})$ parameter. In
this way, one can see an auxiliary  metric function
$h_{\gamma\delta}$ to be proportional to a metric tensor
$g_{\gamma\delta}$, whence $f(R)$ theory would describe the
dynamical properties of $\Lambda$-dominated epoch. Under these
circumstances, the Levi-Civita connection will become the connection
associated with $h_{\gamma\delta}$ given as follows
 \begin{equation}\label{5a}
\Gamma^\nu_{\gamma\delta}=\frac{1}{2}h^{\nu\alpha}(\partial_\gamma h_{\alpha\delta}+\partial_\delta h_{\alpha\gamma}-\partial_\alpha h_{\gamma\delta}).
\end{equation}
We are now focusing on a metric equation of motion, which would of
the second order. To this end, we find the value of connection from
Eq.(\ref{2}). If the same is used in Eq.(\ref{3}), a single field
equation can be found as under
\begin{eqnarray}\nonumber
&&\frac{1}{f_R}\left(\hat{\nabla}_\gamma\hat{\nabla}_\delta-g_{\gamma\delta}
\hat{\Box}\right)f_R+\frac{1}{2}g_{\gamma\delta}\hat{R}+\frac{\kappa}{f_R}
T_{\gamma\delta}+\frac{1}{2}g_{\gamma\delta}\left(\frac{f}{f_R}-\hat{R}\right)
\\\label{4}
&&+\frac{3}{2f_R^2}\left[\frac{1}{2}g_{\gamma\delta}(\hat{\nabla}
f_R)^2-\hat{\nabla}_\gamma f_R\hat{\nabla}_\delta
f_R\right]-\hat{R}_{\gamma\delta}=0,
\end{eqnarray}
Alternatively, this can be expressed via the Einstein tensor
$\hat{G}_{\gamma\delta}$ as.
\begin{equation}\label{6}
\hat{G}_{\gamma\delta}=\frac{\kappa}{f_R}(T_{\gamma\delta}
+{\mathcal{T}_{\gamma\delta}}),
\end{equation}
where $\hat{\Box}$ is an operator of d'Alembertian, and
\begin{eqnarray*}\nonumber
{\mathcal{T}_{\gamma\delta}}&=&\frac{1}{\kappa}\left(\hat{\nabla}_\gamma\hat{\nabla}_
\delta-g_{\gamma\delta}\hat{\Box}\right)f_R-\frac{f_R}{2\kappa}g_{\gamma\delta}
\left(\hat{R}-\frac{f}{f_R}\right)\\\nonumber
&+&\frac{3}{2{\kappa}f_R}\left[\frac{1}{2}g_{\gamma\delta}(\hat{\nabla}
f_R)^2-\hat{\nabla}_\gamma f_R\hat{\nabla}_\delta f_R\right],
\end{eqnarray*}
It is worthy to mention that here the operator $\hat{\nabla}_\gamma$ describes the covariant derivative for $g_{\gamma\delta}$.

We now model our problem with the following plane symmetric metric
\begin{equation}\label{8}
ds^2=-dt^2+B^2(t,z)(dx^2+dy^2)+C^2(t,z)dz^2.
\end{equation}
We assume that the gravitational source of the above planar metric as seen by an observer having a locally Minkowkain frame (LMF)
resting in a comoving congruences is given by
\begin{equation}\label{9}
T_{\gamma\delta}=\hat{\rho} u_\gamma u_\delta
\end{equation}
where $\hat{\rho}$ is the dust energy density. In this environment, the fluid velocity takes the form
\begin{equation}\label{10}
u^\gamma=(1,0,0,0).
\end{equation}
We assume that our systems has a relativistic matter content in the
presence of electromagnetic field. The role of electric charge can
be analyzed through the following tensor
\begin{equation}\label{8x}
E_{\mu\nu}=\frac{1}{4\pi}(F_{\mu}^{~\gamma}F_{\gamma\nu}-\frac{1}{4}F^{\gamma\delta}F^{\gamma\delta}g_{\mu\nu}),
\end{equation}
where $F_{\mu\nu}$ is a Maxwell tensor that can be expressed through
4-potential $(\psi_{\mu})$ as
$F_{\mu\nu}=-\psi_{\mu,\nu}+\psi_{\nu,\mu}$. It satisfy the
equations of motion for the charged medium given follows
\begin{equation}\nonumber
F^{\mu\nu}_{;\nu}=\mathcal{M}J^{\mu},\quad F_{[\mu\nu;\gamma]}=0,
\end{equation}
where $J^{\mu}$ is the 4-current with its magnetic permeability
$\mathcal{M}$. For the current reference frame, we take
$$\Psi_{\nu}=\psi\delta_{\nu}^{0},\quad J^{\mu}=\Gamma u^{\mu},$$
where $\Gamma(t,r)$ stands for the density associated with the
charged medium. In an environment of the non-tilted planar
configuration, we get
\begin{align}\label{9x}
&\frac{\partial^2\psi}{\partial r^2}-\frac{\partial\psi}{\partial
r}\left(\frac{C'}{C}-\frac{2B'}{B}-\frac{2f'_{R}}{f_{R}}\right)=4\pi\Gamma
B^2,\\\label{10x} &\frac{\partial^2\psi}{\partial r\partial
t}-\frac{\partial\psi}{\partial
r}\left(\frac{\dot{B}}{B}-\frac{2\dot{C}}{C}-\frac{2\dot{f_{R}}}{f_{R}}\right)=0,
\end{align}
where over dots and over primes are the notations for the time and
radial partial differentiations, respectively. Equation (\ref{9x})
provides
$$\psi'=\frac{C\hat{\mathfrak{s}}}{B^2f_{R}^2},$$ having
\begin{equation}\nonumber
\hat{\mathfrak{s}}=4\pi \int_{0}^{r} \Gamma CB^2f^2_{R} dr,
\end{equation}
to be the total amount of charge within the manifold of the dust
cloud. From Eq.(\ref{8x}), we obtain
\begin{equation}\nonumber
E_{00}=\frac{\hat{\mathfrak{s}}^2}{8\pi B^4f^4_{R}}, ~E_{01}=0,
~E_{11}=\frac{-C^2\hat{\mathfrak{s}}^2}{8\pi B^4f^4_{R}},
~E_{22}=\frac{~E_{33}}{\sin^2\theta}=\frac{\hat{\mathfrak{s}}^2}{8\pi
B^2f^4_{R}}.\end{equation} The application of a Lorentz boost from
the dust source accompanying LMF to an auxiliary LMF associated with
a fluid radial velocity $(\chi)$ provides a radiating and locally
anisotropic (having pressure components $P_z$ and $P_\bot$) stress
energy tensor as follows
\begin{equation}\label{12}
T_{\gamma\delta}=({\rho}+{P}_{\bot})U_{\gamma}U_{\delta}+\epsilon
l_\gamma l_\delta-{P}_{\bot}g_{\gamma\delta}+{q}_{\gamma}U_\delta
+({P}_z-{P}_{\bot})S_\gamma S_\delta +{q}_{\delta}U_\gamma,
\end{equation}
where
\begin{equation}\label{11}
U^\gamma=\left(\frac{1}{\sqrt{1-\chi^2}},\frac{\chi}{B\sqrt{1-\chi^2}},0,0\right)
\end{equation}
describes the vector field for the tilted reference frame. In
Eq.\eqref{12}, the structural variables $\epsilon,~{\rho}$ explain
radiation and energy densities, respectively, while the vector
corresponding to the heat conduction is given by ${q}_\gamma$. We
further assume that an observer resting in a tilted congruences
spotted the radiation transmissions through the gravitating source
in both diffusion and streaming out approximations. Therefore, we
have taken two distinct factors of dissipation in Eq.\eqref{12}. We
define the vector fields $S^\eta,~q^\eta$ and $l^\eta$ having tilted
backgrounds as follows
\begin{equation}\label{15}
S^\gamma=\left(\frac{\chi}{\sqrt{1-\chi^2}},\frac{1}{B\sqrt{1-\chi^2}},0,0\right),\quad
q^\gamma=qS^\gamma,
\end{equation}
\begin{equation}\label{14}
l^\gamma=\left(\frac{1+\chi}{\sqrt{1-\chi^2}},\frac{1+\chi}{B\sqrt{1-\chi^2}},0,0\right),
\end{equation}
which obey the following constraints
\begin{equation*}
U^{\gamma}U_{\gamma}=-1=l_\gamma U^\gamma,\quad
S^{\gamma}S_{\gamma}=1=l_\gamma S^\gamma,\quad
l^{\gamma}l_{\gamma}=0=S^\gamma U_\gamma=U^\gamma q_\gamma.
\end{equation*}
The equations of motion for the planar tilted observer provide
\begin{eqnarray}\label{16x} \frac{\partial^2\psi}{\partial
r^2}-\frac{\partial\psi}{\partial
r}\left(\frac{C'}{C}-\frac{2B'}{B}-\frac{2f'_{R}}{f_{R}}\right)&=&\frac{4\pi\Gamma
C^2}{\sqrt{1-\chi^2}},\\\label{17x} \frac{\partial^2\psi}{\partial
r\partial t}-\frac{\partial\psi}{\partial
r}\left(\frac{\dot{C}}{C}-\frac{2\dot{B}}{B}-\frac{2\dot{f_{R}}}{f_{R}}\right)&=&-\frac{4\pi\Gamma\omega
C}{\sqrt{1-\chi^2}}.
\end{eqnarray}
On solving Eq.(\ref{16x}), we obtain
$$\psi'=\frac{C\tilde{\mathfrak{s}}}{B^2f^2_{R}},$$ in which $\tilde{\mathfrak{s}}$ describes the total amount of charge
in the non-comoving relativistic matter for the planar geometry. Its
expression is found as under
\begin{equation}\nonumber
{\tilde{\mathfrak{s}}}=4\pi \int_{0}^{r}\frac{\Gamma C
B^2f^2_{R}}{\sqrt{1-\chi^2}} dr.
\end{equation}
From Eq.(\ref{8x}), we get
\begin{equation}\nonumber
E_{00}=\frac{\tilde{\mathfrak{s}}^2}{8\pi B^4f^4_{R}}, ~E_{01}=0,
~E_{11}=\frac{-C^2\tilde{\mathfrak{s}}^2}{8\pi B^4f^4_{R}},
~E_{22}=\frac{~E_{33}}{\sin^2\theta}=\frac{\tilde{\mathfrak{s}}^2}{8\pi
B^2f^4_{R}}.\end{equation} These are the non-vanishing components of
the electromagnetic energy momentum tensor for the planar charged
medium whose observations are noticed by a tilted observer.

\subsection{Relations between Comoving and Non-comoving Reference Frames with an Electromagnetic field}

From Eq.\eqref{6}, the corresponding Palatini $f(R)$ field equations for the charged anisotropic plane symmetric systems can be written as follows
\begin{align}\nonumber
G_{00}&=\frac{\kappa}{f_R}\left[\frac{1}{1-\chi^2}\left\{\tilde{P}_r\chi^2+\tilde{\rho}+2\tilde{q}\chi\right\}+\frac{1}{\kappa}
\left\{-\left(\frac{f}{R}-f_R\right)\frac{R}{2}-\dot{f_R}\left(\frac{\dot{C}}{C}
+\frac{9}{4}\right. \right.\right.\\\label{16}
&\times\left.\left.\left.\frac{\dot{f_R}}{f_R}+\frac{2\dot{B}}{B}\right)-\left(\frac{C'}{C}-\frac{2B'}{B}
+\frac{1}{4}\frac{f'_R}{f_R}\right)\frac{f'_R}{C^2}
+\frac{f''_R}{C^2}\right\}+\frac{\tilde{\mathfrak{s}}^2}{8\pi
B^4f_R^4}\right],\\\nonumber
\frac{G_{11}}{B^2}&=\frac{\kappa}{f_R}\left[P_\bot+\frac{1}{\kappa}
\left\{\left(\frac{f}{R}-f_R\right)\frac{R}{2}-\frac{f''_R}{C^2}+\ddot{f_R}
+\left(\frac{4\dot{C}}{C}-\frac{\dot{f_R}}{f_R}+\frac{4\dot{B}}{B}\right)\frac{\dot{f_R}}{4}\right.\right.\\\label{18}
&\left.\left.+\left(\frac{4C'}{C}+\frac{f'_R}{f_R}-\frac{4B'}{B}\right)\frac{f_R'}{4C^2}\right\}+\frac{\tilde{\mathfrak{s}}^2}{8\pi
B^4f_R^4}\right],\\\label{19}
G_{03}&=\frac{\kappa}{f_R}\left[-\frac{C}{(1-\chi^2)}\left\{(\tilde{P}_z+\tilde{\rho})\chi
+(1+\chi^2)\tilde{q}\right\}
+\frac{\dot{f_R}}{\kappa}\left\{\frac{\dot{f'_R}}{\dot{f_R}}-\frac{5}{2}\frac{f_R'}{f_R}
-\frac{f'_R\dot{C}}{C\dot{f_R}}\right\}\right],\\\nonumber
\frac{G_{33}}{C^2}&=\frac{\kappa}{f_R}\left[\frac{1}{(1-\chi^2)}\left\{\tilde{P}_z+\tilde{\rho}\chi^2
+2\tilde{q}\chi\right\}+\frac{1}{\kappa}
\left\{\left(\frac{f}{R}-f_R\right)\frac{R}{2}-\frac{f'_R}{4C^2}
\left({9}\frac{f'_R}{f_R}\right.\right.\right.\\\label{17}
&+\left.\left.\left.\frac{8B'}{B}\right)+\ddot{f_R}-
\left(\frac{\dot{f_R}}{f_R}-\frac{8\dot{B}}{B}\right)\frac{\dot{f_R}}{4}\right\}-\frac{\tilde{\mathfrak{s}}^2}{8\pi
B^4f_R^4}\right],
\end{align}
where the tilde over the quantities describes that the corresponding quantities are evaluated after adding
$\epsilon$ in them, for instance, $\tilde{y}\equiv y+\epsilon$, while the values
of Einstein tensor $G_{ii}'s$ can be found at \cite{herrera2011tilted}.

In the analysis of stellar systems with tilted or non-tilted
congruences, the selection of $f(R)$ models is of significant
importance. Here, we would like to study the influence of $f(R)$
gravity on bringing the importance of congruences of the observers
in the explanations of physical phenomena under certain feasible
theoretical backgrounds. It is worthy to stress that the use of
quadratic Ricci curvature corrections is among the viable attempts
to renormalize GR as an alternative to the most conventional
gravitational theory. The inclusion of these corrections in the EHA
occupies key relevance in the field of theoretical cosmology. These
could assist one to analyze the dynamical features of galaxies,
cluster, their evolutions as well as compact objects in an arena of
self-consistent inflationary environment. Such terms are induced as
an approximation for the study of DE models.

There is growing evidence that the Universe at the present epoch is
undergoing an exponential expansion process. The preferred reason
for this phenomenon is that a sort of dark energy is actually
dominating the Universe. It is worthy to stress that, the dynamics
of the early universe can be well discussed through the higher
derivative corrections containing the positive curvature terms. The
terms with negative powers of curvature could act as a gravitational
alternative to allow the interstellar acceleration to become
consistent with DE \cite{nojiri2007introduction}. In order to study
such epochs in a unified way, we take, with the non-tilted
gravitational source (\ref{9}), the following model
\begin{equation}\label{20}
f(R)=R+{\psi}R^{2},
\end{equation}
in which $\psi$ is a constant \cite{starobinsky1980new} which may
equals $\frac{1}{6M^2}$ (with $M$ as $2.7\times10^{-12}GeV$ along
with $\psi\leq2.3\times10^{22}Ge/V^2$) for the study of the exotic
matter, that appear precisely at the present times. We take the
$f(R)$ model associated with the non-comoving reference frame
(\ref{12}) as follows \cite{PhysRevD.75.127502}
\begin{equation}\label{21}
f(R)=R+\rho_1\frac{\delta^4}{R},
\end{equation}
where $\delta\in(0,\infty)$ and $\rho_1=+1$. The choice
$\delta^{-1}\thicksim(10^{33}eV)^{-1}\thicksim10^{18}\textrm{sec}$
may helps to understand the evolution of our cosmos at present
times.

The gravitating sources connected with the both types of tilted and
non-tilted systems are producing the same kind of geometric
structure. Therefore, the connections between matter variables along
with the effective curvature terms can be calculated. The same
relations are being found as under
\begin{align}\nonumber
&\frac{\kappa\hat{\rho}}{1+2\psi
R}+\frac{\kappa\tilde{\mathfrak{s}}^2}{8\pi B^4(1+2\psi
R)^5}+\frac{\psi R^2}{2(1+2\psi
R)}+\frac{R\delta^2}{R^2-\delta^4}=\frac{\kappa\tilde{\mathfrak{s}}^2R^{10}}{8\pi
B^4(R^2-\delta^4)^5}\\\label{27} &\frac{\kappa
R^2(\tilde{\rho}+\tilde{P}_r\chi^2+2\tilde{q}\chi)}{(R^2-\delta^2)
(1-\chi^2)},\\\label{28}
&(\tilde{\rho}+\tilde{P}_r)\chi+\tilde{q}(1+\chi^2)=0,\\\nonumber
&\frac{\kappa
R^2(\tilde{\rho}\chi^2+\tilde{P}_r+2\tilde{q}\chi)}{(R^2-\delta^2)(1-\chi^2)}+\frac{\kappa\hat{\mathfrak{s}}^2}{8\pi
B^4(1+2\alpha R)^5}=\frac{-\psi R^2}{2(1+2\psi
R)}-\frac{\kappa\tilde{\mathfrak{s}}^2R^{10}}{8\pi
B^4(R^2-\delta^4)^5}\\\label{29}&-\frac{R\delta^2}{R^2-\delta^2},\\\label{30}
&P_\bot=\frac{-\psi(R^2-\delta^4)}{2(1+2\psi
R)\kappa}-\frac{\tilde{\mathfrak{s}}^2R^8}{8\pi
B^4(R^2-\delta^4)^4}-\frac{\delta^4}{R\kappa}+\frac{\hat{\mathfrak{s}}^2(R^2-\delta^4)}{8\pi
B^4R^2 (1+2\alpha R)^5}.
\end{align}
These relations are connecting gravitating variables of
non-interacting particles found in inflation field with the fluid
variables of non-ideal radiating locally anisotropic matter resting
in the cosmic speed up epoch. The electromagnetic field and Palatini
$f(R)$ terms are making the role of $P_\bot$ to be non-zero that was
zero in the background of Einstein gravity \cite{herrera2011tilted}.
This shows that these two forces are producing special effects
(equal to tangential pressure of the non-tilted gravitating source)
on the dynamics of the system. Some more relations among the
variables of the charged fluid configurations of the above mentioned
systems can be found by making use of Eqs.(\ref{9}) and (\ref{12})
as
\begin{align}\label{22}
\epsilon&=\frac{(R^2-\delta^4)}{1+2\psi
R}\left\{\frac{\hat{\rho}}{R^2(1-\chi^2)}+\frac{\hat{\mathfrak{s}}^2}{8\pi
B^4R^2(1+2\psi
R)^4(1-\chi^2)}+\frac{\psi}{2\kappa}\right\}+\frac{\delta^4}{R\kappa}-\rho,\\\nonumber
P_r&=\rho-\frac{(R^2-\delta^4)\hat{\rho}}{R^2(1+2\psi
R)}-\frac{\hat{\mathfrak{s}}^2(R^2-\delta^4)}{8\pi B^4R^2(1+2\psi
R)^5}-\frac{(R^2-\delta^4)\psi}{\kappa(1+2\psi
R)}-\frac{2\delta^4}{R\kappa}\\\label{23}
&+\frac{\tilde{\mathfrak{s}}^2R^8}{8\pi
B^4(R^2-\delta^4)^4},\\\label{24}
q&=\frac{-\hat{\rho}(R^2-\delta^4)\chi}{R^2(1+2\psi
R)(1-\chi^2)}-\frac{\hat{\mathfrak{s}}^2(R^2-\delta^4)\chi}{8\pi
B^4R^2(1+2\psi
R)^5(1-\chi^2)}+\frac{\tilde{\mathfrak{s}}^2R^8}{(R^2-\delta^4)^4}\\\nonumber
&\times\frac{\chi}{8\pi B^4(1-\chi^2)}-\epsilon,\\\nonumber
&=\rho-\frac{\hat{\rho}(R^2-\delta^4)}{R^2(1-\chi)(1+2\psi
R)}-\frac{\hat{\mathfrak{s}}^2(R^2-\delta^4)}{8\pi B^4R^2(1+2\psi
R)^5(1-\chi)}-\frac{\delta^4}{R\kappa}\\\label{25}
&+\frac{\tilde{\mathfrak{s}}^2R^8}{8\pi
B^4(R^2-\delta^4)^4(1-\chi)}-\frac{\psi(R^2-\delta^4)}{2\kappa(1+2\chi
R)},\\\label{26}
&=\frac{P_r-\rho\chi}{(1-\chi)}+\frac{\psi(R^2-\delta^4)(1+\chi)}{2\kappa(1+2\psi
R)(1-\chi)}+\frac{\delta^4(1+\chi)}{R\kappa(1-\chi)}.
\end{align}
It is clear from Eq.\eqref{24} that the effects of energy
dissipation stemming from the streaming out and diffusion
approximations could only be seen if the tilted system is able to
retain its four velocity. On the contrary the conditions
$\epsilon=q=0$, would keep the value of tilted four velocity to be
zero, even though we are observing this analysis in the background
of Maxwell-Palatini $f(R)$ theory. Thus, one can state that the
observations of heat dissipation from the charged relativistic
matter is only possible if a system is able to retain the
configurations of non-comoving reference frame. Taking this into
consideration, we now provide few relationships under some
particular constraints.

\subsubsection{$\epsilon\neq0$ in the background of Zero Diffusion Approximations}

In this scenario, the structural quantities of plane symmetric model
in the presence of electromagnetic field in view of
Eqs.(\ref{22})-(\ref{26}) provide
\begin{align}\label{34}
P_r&=\rho\chi-\frac{\psi(R^2-\delta^4)(1+\chi)}{2\kappa(1+2\psi
R)}-\frac{\delta^4(1+\chi)}{\kappa R},\\\nonumber
\rho&=\frac{(R^2-\delta^4)}{(1+2\psi
R)}\left\{\frac{\hat{\rho}}{R^2(1-\chi)}+\frac{\psi}{2\kappa}+\frac{\hat{\mathfrak{s}}^2}{8\pi
B^4R^2(1+2\psi R)^4(1-\chi)}\right\}\\\label{35}
&-\frac{\tilde{\mathfrak{s}}^2R^8}{8\pi
B^4(R^2-\delta^4)^4(1-\chi)}+\frac{\delta^4}{R\kappa},\\\nonumber
\epsilon&=\frac{-\hat{\rho}(R^2-\delta^4)\chi}{R^2(1+2\psi
R)(1-\chi^2)}-\frac{\hat{\mathfrak{s}}^2(R^2-\delta^4)\chi}{8\pi
B^4R^2(1+2\psi R)^5(1-\chi^2)}\\\label{36}
&+\frac{\tilde{\mathfrak{s}}^2R^8\chi}{8\pi
B^4(R^2-\delta^4)^4(1-\chi^2)}.
\end{align}

\subsubsection{$q\neq0$ in an Environment of Zero Streaming Out Approximations}

In this end, the planar charged matter variables are found to be
related through Eqs.(\ref{22})-(\ref{26}) as follows
\begin{align}\label{31}
P_r&=\rho\chi^2-\frac{\psi(R^2-\delta^4)(1+\chi^2)}{2\kappa(1+2\psi
R)}-\frac{\delta^4(1+\chi^2)}{\kappa R},\\\nonumber
\rho&=\frac{(R^2-\delta^4)}{(1+2\psi
R)}\left\{\frac{\hat{\rho}}{R^2(1-\chi^2)}+\frac{\hat{\mathfrak{s}}^2}{8\pi
B^4R^2(1+2\psi
R)^4(1-\chi^2)}+\frac{\psi}{2\kappa}\right\}\\\label{32}
&-\frac{\tilde{\mathfrak{s}}^2R^8}{8\pi
B^4(R^2-\delta^4)^4(1-\chi^2)}+\frac{\delta^4}{R\kappa},\\\nonumber
q&=\frac{-\hat{\rho}(R^2-\delta^4)\chi}{R^2(1+2\psi
R)(1-\chi^2)}-\frac{\hat{\mathfrak{s}}^2}{8\pi B^4R^2(1+2\psi
R)^5(1-\chi^2)}+\frac{\tilde{\mathfrak{s}}^2}{8\pi B^4}\\\label{33}
&\times \frac{R^8\chi}{(R^2-\delta^4)^4(1-\chi^2)}.
\end{align}

\subsubsection{$P_r=0$}

It becomes from Eqs.(\ref{22})-(\ref{26}) as
\begin{align}\nonumber
\rho&=\frac{\hat{\rho}(R^2-\delta^4)}{R^2(1+2\psi
R)}+\frac{\hat{\mathfrak{s}}^2(R^2-\delta^4)}{8\pi B^4R^2(1+2\psi
R)^5}+\frac{\psi(R^2-\delta^4)}{\kappa(1+2\psi R)}\\\label{37}
&-\frac{\tilde{\mathfrak{s}}^2R^8}{8\pi
B^4(R^2-\delta^4)^4}+\frac{2\delta^4}{\kappa R},\\\label{38}
q&=\frac{-\rho\chi}{1-\chi}+\frac{\psi(R^2-\delta^4)(1+\chi)}{2\kappa
(1+2\psi
R)(1-\chi)}+\frac{\delta^4(1+\chi)}{R\kappa(1-\chi)},\\\label{39}
\epsilon&=\frac{\rho\chi^2}{(1-\chi^2)}-\frac{\psi(R^2-\delta^4)(1+\chi^2)}{2\kappa
(1+2\psi R)(1-\chi^2)}-\frac{\delta^4(1+\chi^2)}{R\kappa(1-\chi^2)}.
\end{align}

\section{Kinematical Quantities}

This section deals with the computation of kinematical quantities of
the charged non-static plane symmetric model in $f(R)$ theory with
Palatini formalism. The investigation of such terms with $f(R)$
model \eqref{21} could help us to understand physical features of
planar irrotational locally anisotropic radiating interiors. We
shall evaluate such properties one by one as follows.

The generic formula to calculate the four acceleration of the
collapsing matter configurations can be specified as
\begin{align}\label{40}
a^\gamma&=U^{\gamma}_{~;\delta}U^\delta.
\end{align}
This with the help of $f(R)$ extra curvature corrections is found as
under
\begin{align}\label{41}
a^\gamma&=aS^\gamma-g^{\gamma\gamma}\frac{\partial_\gamma
f_R}{2f_R},
\end{align}
where
\begin{align}\label{42}
a=\frac{1}{\sqrt[3]{1-\chi^2}}\left[\dot{\chi}+\frac{\chi\chi'}{C}+(1-\chi^2)\left(\frac{\chi\dot{C}}{C}
+\frac{2\chi\delta^4R^{-3}\dot{R}}{(1-\delta^4R^{-2})}
+\frac{2\delta^4R^{-3}R'}{C(1-\delta^4R^{-2})}\right)\right].
\end{align}
Since we have figured out the non-zero contribution of the
$4$-acceleration, therefore, the observer residing in the
non-comoving framework identified the non-geodesic nature of the
plane geometric structure. This happened due to the inclusion of
Palatini $f(R)$ gravity in an environment of tilted congruences.
From Eq.\eqref{42}, it can be noticed that the radial velocity
$\chi$ and the modifications in Einstein gravity implied by Palatini
$f(R)$ theory force the process to allow non-geodesic motion of the
test particles.

The scalar for the describing expansion is
\begin{align}\label{43}
\Theta=U^\mu_{~;\mu},
\end{align}
which for the relativistic charged planar system is found as follows
\begin{align}\nonumber
\Theta&=\frac{1}{\sqrt[3]{1-\chi^2}}\left[\chi\dot{\chi}+\frac{\chi'}{C}+(1-\chi^2)\left(\frac{\dot{C}}{C}
+4\frac{\delta^4R^{-3}\dot{R}}{(1-\delta^4R^{-2})}
+\frac{2\dot{B}}{B}+\frac{2\chi B'}{BC}\right.\right.\\\label{44}
&\left.\left.+\frac{2\chi\delta^4R^{-3}R'}{C(1-\delta^4R^{-2})}\right)\right].
\end{align}
The condition $\Theta=0$, would reduce the above equation as follows
\begin{align}\label{44n}
\frac{\delta^4\dot{R}}{R^{3}}=\frac{(1-\delta^4R^{-2})}{4(\chi^2-1)}\left\{\chi\dot{\chi}+\frac{\chi'}{C}+(1-\chi^2)\left(\frac{\dot{C}}{C}
+\frac{2\dot{B}}{B}+\frac{2\chi B'}{BC} +\frac{\chi
2\delta^4R^{-3}R'}{C(1-\delta^4R^{-2})}\right)\right\}.
\end{align}
This equation would help to understand those systems who are having
Minkowkian core.

Here we are dealing with the relativistic model consisting of a
fluid with shear. Therefore, the tensorial quantity for describing
shear fluid is described through projection tensor
$h_{\gamma\delta}$ by
\begin{align}\label{45}
\sigma_{\gamma\delta}=U_{(\gamma;\delta)}+a_{(\gamma}U_{\delta)}-\frac{1}{3}\Theta
h_{\gamma\delta},
\end{align}
whose non-zero values are found as under
\begin{align}\label{46}
\sigma_{00}&=\frac{2\chi^2}{3(1-\chi^2)}\left[\sigma+\frac{3\delta^4R^{-3}\dot{R}}{\chi^2(1-\delta^4R^{-2})\sqrt{1-\chi^2}}
+\frac{3\delta^4R^{-3}R'}{2C(1-\delta^4R^{-2})\chi}\sqrt{1-\chi^2}\right],\\\label{47}
\sigma_{11}&=\frac{2C^2}{3(1-\chi^2)}\left[\sigma+\sqrt{1-\chi^2}
\left\{\frac{3\delta^4R^{-3}\dot{R}}{2(1-\delta^4R^{-2})}-\frac{3\chi\delta^4R^{-3}R'}{Cf_R}\right\}\right],\\\label{48}
\sigma_{22}&=\frac{-B^2}{3}\left[\sigma+\frac{1}{\sqrt{1-\chi^2}}\left\{\frac{3\delta^4R^{-3}\dot{R}}{(1-\delta^4R^{-2})}-\frac{\chi
\delta^4R^{-3}R'}{B(1-\delta^4R^{-2})}\right\}\right],
\end{align}
where
\begin{align}\nonumber
\sigma&=\frac{1}{\sqrt[3]{1-\chi^2}}\left[\chi\dot{\chi}+\frac{\chi'}{C}+(1-\chi^2)
\left\{\frac{\dot{C}}{C}-\frac{2\delta^4R^{-3}\dot{R}}{(1-\delta^4R^{-2})}
-\frac{\dot{B}}{B}-\frac{\chi B'}{BC}\right.\right.\\\label{49}
&\left.\left.
+\frac{2\chi\delta^4R^{-3}R'}{C(1-\delta^4R^{-2})}\right\}\right].
\end{align}
On making $\chi=0$, the value of the shear scalar $\sigma$ reduces
to
\begin{align}\label{49a}
\sigma&=\frac{\dot{C}}{C}-\frac{2\delta^4R^{-3}\dot{R}}{(1-\delta^4R^{-2})}
-\frac{\dot{B}}{B}
\end{align}
that clearly indicates the inclusion of $\delta$ terms in the
analysis. The condition $\sigma=0$ yields the following components
of the shear tensor
\begin{align}\label{46a}
\sigma_{00}&=\frac{2\chi^2\delta^4}{3(1-\chi^2)(1-\delta^4R^{-2})}\left[\frac{3R^{-3}\dot{R}}{\chi^2\sqrt{1-\chi^2}}
+\frac{3R^{-3}R'}{2C\chi}\sqrt{1-\chi^2}\right],\\\label{47a}
\sigma_{11}&=\frac{2C^2\delta^4}{3(1-\chi^2)(1-\delta^4R^{-2})}\left[\sqrt{1-\chi^2}\left(\frac{3\dot{R}}{2R^{3}}-\frac{3\chi
R^{-3}R'}{C}\right)\right],\\\label{48a}
\sigma_{22}&=\frac{-B^2\delta^4}{3(1-\delta^4R^{-2})}\left[\frac{1}{\sqrt{1-\chi^2}}\left(\frac{3\dot{R}}{R^{3}}-\frac{2\chi
R^{-3}R'}{2C}\right)\right],
\end{align}
which formulate that Palatini $\delta$ corrections explicitly tend
to regulate the shear-free phases of the locally anisotropic
dissipative planar metric.

\section{Equations of Motion}

The purpose of this section is to compute dynamical, collapse and
Weyl scalar equations. We shall calculate dynamical and collapse
equations from the contracted form of modified Bianchi identities,
while the Weyl scalar equation will be computed through the
procedure presented firstly by Ellis \cite{ellis2012relativistic}and
then by Herrera \cite{herrera2011physical}. These mathematical
expressions will help us to explain the appearance of irregularities
in the charged planar celestial population which is assumed to be
initially homogeneous in nature. The equations for the description
of radial and temporal fluctuations in the collapsing interiors in
the presence of Maxwell-Palatini $f(R)$ corrections can be
calculated from the following identities
\begin{eqnarray}\nonumber
Y^\gamma_{~\delta_;\gamma}=0,~~\textrm{where}~~
Y^\gamma_{~\delta}=E^\gamma_{~\delta}+T^\gamma_{~\delta}+\mathcal{T}^\gamma_{~\delta}.
\end{eqnarray}
Now, we define two operators $\dag$ and $*$ as follows
\begin{align}\nonumber
y^\dag=y_{,\nu}S^\nu,\quad y^*=y_{,\nu}U^\nu.
\end{align}
After using Eqs.\eqref{16}-\eqref{19}, \eqref{42}, \eqref{44} and
\eqref{49}, we get
\begin{align}\nonumber
&\tilde{\rho}^*+\tilde{\rho}\Theta+\tilde{q}^\dag+\tilde{q}\left\{\chi\Theta
+\frac{\sqrt{1-\chi^2}}{C}\left(\frac{2B'}{B}+\frac{f'_R}{f_R}\right)
+\frac{2\dot{\chi}}{\sqrt{1-\chi^2}}\right\}+\frac{\tilde{\rho}f_R^*}{f_R}\\\nonumber
&+\frac{\tilde{q}f_R^\dag}{f_R}+\frac{\chi
P_\bot'}{C\sqrt{1-\chi^2}}+\frac{\tilde{\mathfrak{s}}^2\dot{f_{R}}}{4\pi
B^4f_{R}^5\sqrt{1-\chi^2}}+P_\bot\left(\Theta+\frac{\dot{f_R}}{f_R\sqrt{1-\chi^2}}+\frac{\chi
f_R'}{f_R\sqrt{1-\chi^2}}\right)\\\label{50}
&+\frac{\mathcal{K}_0}{\sqrt{1-\chi^2}}=0,\\\nonumber
&\tilde{P_z}^\dag+a(\tilde{\rho}+\tilde{P_z})+\frac{2\tilde{q}}{3}\left[2\Theta+\sigma-3\chi(\ln
B)^\dag\right]+\tilde{q}^*+\frac{\chi
f_R^*}{f_R}(\tilde{\rho}+P_\bot)-\tilde{q}\sqrt{1-\chi^2}\\\nonumber
&\times
\left(\frac{\dot{C}}{C}+\frac{2\dot{B}}{B}\right)+\frac{1}{f_R\sqrt{1-\chi^2}}\left(\tilde{q}\chi^2\dot{f_R}
-\frac{\tilde{\rho}f'_R}{C}-\frac{P_\bot
f_R'}{C}\right)-\frac{\tilde{\mathfrak{s}}\tilde{\mathfrak{s}'}}{4\pi
B^4f_{R}^{4}}+\frac{\tilde{\mathfrak{s}}^{2}f'_{R}}{4\pi
B^4f_{R}^5}\\\label{51}
&-\frac{\chi^2P_\bot'}{C\sqrt{1-\chi^2}}+\frac{\chi}{\sqrt{1-\chi^2}}
[\dot{\tilde{\mu}}+(\chi\tilde{q}\dot{)}]+\frac{\mathcal{K}_1\sqrt{1-\chi^2}}{C}-\sqrt{1-\chi^2}(P_{\bot}\chi\dot{)}=0,
\end{align}
where $\mathcal{K}_i$'s describe the part of the role emerging from
the Palatini $f(R)$ gravity. These are mentioned in an Appendix.
Upon substitution of Eq.\eqref{44n}, Eq.\eqref{51} with
expansion-free condition provides
\begin{align}\nonumber
&\tilde{P_z}^\dag+a(\tilde{\rho}+\tilde{P_z})+\frac{2\tilde{q}}{3}\left[\sigma-3\chi(\ln
B)^\dag\right]+\tilde{q}^*+\frac{\chi
f_R^*}{f_R}(\tilde{\rho}+P_\bot)-\tilde{q}\sqrt{1-\chi^2}\\\nonumber
&\times
\left(\frac{\dot{C}}{C}+\frac{2\dot{B}}{B}\right)+\frac{1}{f_R\sqrt{1-\chi^2}}\left(\tilde{q}\chi^2\Psi
{f_R} -\frac{\tilde{\rho}f'_R}{C}-\frac{P_\bot
f_R'}{C}\right)-\frac{\tilde{\mathfrak{s}}\tilde{\mathfrak{s}'}}{4\pi
B^4f_{R}^{4}}+\frac{\tilde{\mathfrak{s}}^{2}f'_{R}}{4\pi
B^4f_{R}^5}\\\label{51n}
&-\frac{\chi^2P_\bot'}{C\sqrt{1-\chi^2}}+\frac{\chi}{\sqrt{1-\chi^2}}
[\dot{\tilde{\mu}}+(\chi\tilde{q}\dot{)}]-\sqrt{1-\chi^2}(P_{\bot}\chi\dot{)}+\frac{\mathcal{K}_2\sqrt{1-\chi^2}}{C}=0,
\end{align}
where
\begin{align}\label{44nn}
\Psi=\frac{1}{2(\chi^2-1)}\left\{\chi\dot{\chi}+\frac{\chi'}{C}+(1-\chi^2)\left(\frac{\dot{C}}{C}
+\frac{2\dot{B}}{B}+\frac{2\chi B'}{BC} +\frac{\chi
f'_R}{Cf_R}\right)\right\},
\end{align}
while $\mathcal{K}_2$ is being calculated by putting zero expansion
condition in the expression of $\mathcal{K}_1$. This equation
describes the radial variations on the matter variables of those
relativistic objects who are able to create empty core during
evolution.

A relativistic self-gravitating fluid may reach at the collapsing
phase after experiencing energy density irregularities in its energy
density. Based about how much massive a celestial structure is, the
phenomenon of gravitational collapse may produce various types of
compact objects, like black hole, neutron star or white dwarf.
Therefore, the research for the elements of inhomogeneity in the
initially regular spacetimes is of considerable importance. Here, in
an atmosphere of tilted anisotropic configuration, we determine
those variables that are involved in producing disturbances on the
homogeneous energy density of the planar relativistic interiors
evolving an environment of electromagnetic field. A mathematical
combinations of Weyl scalar and effective form of the charged plane
symmetric fluid variables can be devised to obtain such factor
\cite{yousaf2019role}. For non-comoving congruences, it follows that
\begin{align}\nonumber
&\left[\mathcal{E}-\frac{\kappa}{2f_R}\left(\tilde{\rho}-\tilde{P}_z+{P}_\bot+\mathcal{T}_{00}
-\frac{\mathcal{T}_{11}}{C^2}+\frac{\mathcal{T}_{33}}{B^2}-\frac{\tilde{\mathfrak{s}}^2}{8\pi
B^4f_R^4}\right)\right]'=-\frac{3B'}{B}\left[\mathcal{E}
+\frac{\kappa}{2f_R}\right.\\\nonumber
&\left.\times\left(\tilde{P}_z-{P}_\bot+\frac{\mathcal{T}_{11}}{C^2}+\frac{\tilde{\mathfrak{s}}^2}{8\pi
B^4f_R^4}-\frac{\mathcal{T}_{33}}{B^2}\right)\right]
+\frac{3\kappa\dot{B}}{2Bf_R}\left\{\frac{(\tilde{\rho}+\tilde{P}_z)\chi+\tilde{q}(1+\chi^2)}{(1-\chi^2)}\right.\\\label{52}
&\left. -\frac{\mathcal{T}_{03}}{C}\right\}.
\end{align}
This is a non-linear partial differential equation. The solution to
this equation is very tough and complicated. However, the analytical
approach for getting solution of the above equation may be done by
considering certain viable explanations. The model that could help
to explain the current cosmic acceleration in the realm of an
inhomogeneous background is LTB. The considerations of tilted and
non-tilted congruences in LTB will not produce hindrances in the
description of inhomogeneity picture. But the situation about the
appearances of irregularities in the energy density is quite
different for the tilted framework with planar LTB-like spacetime.
In the following, we wish to calculate the factors that are
participating in the evolution of the charged planar system from the
homogeneous phase to the inhomogeneous window with Maxwell-Palatini
$f(R)$ corrections.

In this direction, we assume that our plane symmetric relativistic
model develops in a such a way that the entire emission and
absorbtion through the matter distribution is null and thereby
making the system to retain its only isotropic pressure effects.
Therefore, the conditions $P_z=P_\bot=P$ and $q=\epsilon=0$ would
reduce Eq.\eqref{52} to
\begin{align}\label{53}
\mathcal{E}'+\frac{3B'}{B}\mathcal{E}=\frac{\kappa}{2}\left[\frac{1}{f_R}\left(\tilde{\rho}
+\mathcal{T}_{00}-\frac{\mathcal{T}_{11}}{C^2}-\frac{\tilde{\mathfrak{s}}^2}{8\pi
B^4f_R^4}+\frac{\mathcal{T}_{33}}{B^2}\right)\right]'
+\frac{3\kappa\dot{B}}{2Bf_R}\frac{(\tilde{\rho}+\tilde{P})}{(1-\chi^2)}\chi.
\end{align}
It has been well recognized that the scalar that governs the impact
of energy density irregularities of the galactic and inter-galactic
celestial population surface of an object is the Weyl tensor. Here,
we want the same type of variable to be determined for the plane
symmetric spacetime with Palatini $f(R)$ gravity in the presence of
electromagnetic field. After calculating the solution of the above
partial differential equation, we notice that three elements are
forcing the system to persist in the homogenous phase. These are
effective electric charge $\mathfrak{s}$, Palatini $f(R)$
corrections and velocity $\psi$ mediating from vector field of the
non-comoving congruences. These three quantities are striving to
sustain the planar system to be remain in the initial homogeneous
state. The constraints $\psi=0,~f(R)=R$ reduce the above the above
to
\begin{align}\label{53a}
\tilde{\rho}'=0\Leftrightarrow \mathcal{E}=0.
\end{align}
The stability of gravitational collapse as well as the comoving
congruences for the relativistic geometric system has been studied
by many researchers
\cite{Herrera_1997,HERRERA2002157,PhysRevD.70.084004,yousaf2018some,yousaf2020construction}.

\section{Summary}

After recent developments in the study of modern Physics, (CMB and
supernova-type Ia surveys), interest in the study of plane
symmetrical geometry has been established in order to investigate
few mysterious features of our evolutionary universe. This paper is
aimed to study the effects of Maxwell-Palatini $f(R)$ terms on the
stability of tilted frame for the plane symmetric model. We shall
also try to connect inflationary and late time cosmic speed up eras
with the help of tilted and non-tilted observers. After the
rotation-free transformation of a LMF corresponding to
Eq.\eqref{10}, a scenario of tilted congruences with a gravitating
source possessing a certain velocity in the radial direction can be
created. In this respect, we assume that LTB-like plane symmetric
spacetime is filled observer's dependent two different types of
fluid distributions. An observer resting in a comoving congruence
observe that LTB-like plane symmetric model is generated by a dust
fluid, while the anisotropic radiating matter is a gravitating
source of the same geometry as noticed by a titled observer whose
reference frame is moving with a specific radial velocity $\psi$.

We shall explore the Palatini $f(R)$ and Maxwell field equations for
the plane symmetric collapsing model. As we have taken two different
physical interpretation of congruences therefore, it is worth while
to expect that quantities whose calculations are based on
congruences should play an essential fundamental role in our study.
For the irrotational relativistic study, the congruence of an
observer is linked with three kinematical variables. We investigated
the non-zero components of such variables in the Maxwell-$f(R)$
gravity with Palatini formalism.

We also established relationships between field variables in these
subsequent frames with such contexts to see the influence of
observers on the interpretations of planar geometry dynamics. This
has helped us to study cosmological aspects of a non-tilted model
with a quadratic Ricci inflationary model with the tilted late time
cosmic speed up terms in a unified way with an electromagnetic
field. The effects of electric charge and Palatini $f(R)$ terms are
extensively studied.

As shown in Eq.(\ref{41}), a particular form of the 4-acceleration
is evaluated with Palatini $f(R)$ gravity corrections. We have
determined the non-geodesic nature of LTB-like plane symmetric
model. This happened due to the presence of Maxwell-$f(R)$ terms.
The non-zero values of the shear tensor and a scalar associated with
the expansion scalar is calculated. We have also developed a
relation that could help to understand the cavity production within
the matter distribution of tilted framework. This scenario could be
helpful in the study of cosmological voids. Herrera et al.
\cite{herrera2009expansion} stated that the zero expansion condition
can only be applicable to those relativistic fluids who have
pressure anisotropy in their gravitating sources. Therefore, we can
not apply the condition of expansion-free to the matter source
mentioned in Eq.\eqref{10}. Thus the the phenomenon of cavity
evolution is expected to appear in the tilted charged plane
symmetric congruences during the inflationary cosmic eras. Thus the
tilted framework is likely to host structures of cosmological voids.

The assumption of non-comoving frame of reference could play a
significant role in the explanation of many secrets and hidden
facets of our evolving universe. This concept is sufficiently
versatile to accommodate and describe many celestial systems of
various distributions of matter. We have also studied the stability
of regular energy density of the homogeneous relativistic plane
symmetric model in the presence of electromagnetic field. It is
concluded that the presence of effective charge is trying to
increase the stability of homogeneous energy density of the system.
This finally lead us to find the state of conformal flatness within
the charged planar model in Palatini $f(R)$ gravity. This result has
been obtained after computing the modified version of Ellis
equations from the second Bianchi identity of the charged plane
symmetric metric.

\begin{figure} \centering
\epsfig{file=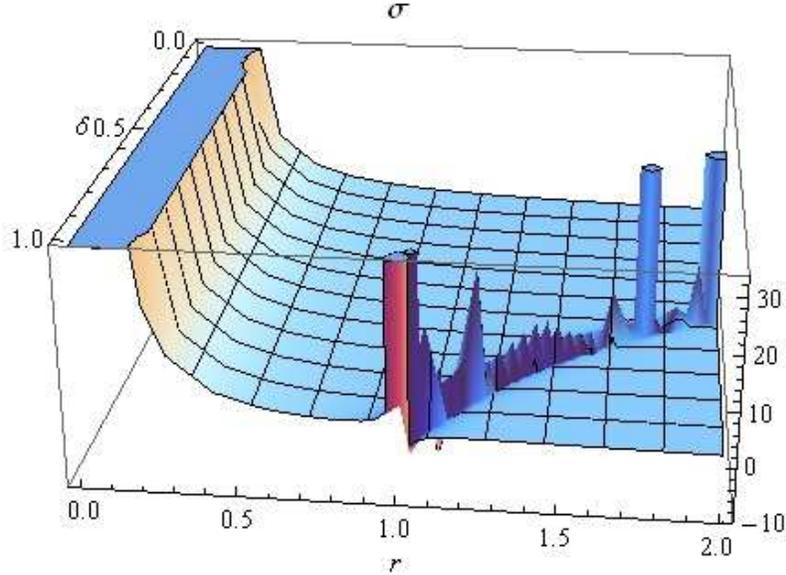,width=0.8\linewidth} \caption{Variation in the shear on the existence of stellar objects
with respect to $\delta$ and
$r$.}\label{shear1}
\end{figure}

There are a collection of structures with well recognized physical properties and thermal history
persists besides supernovae, that could provide valuable and compatible tests for modified gravity theories.
The ultimate objective behind this research is to recognize the most significant changes
in the astrophysical predictions of self-gravitating non-rotational stellar structures in contrast with GR
frameworks. We have noticed some very interesting results because of the modification of gravity.
We have noticed that modification of gravity mediated by $f(R)=R+\frac{\delta^4}{R}$ corrections is likely to host shearing stellar models
than that in GR. The stellar models having stronger shearing effects and unit radii are likely to
exist in more abundance in an environment created by $\delta=1$ parameter. The subsistence of the
self-gravitating systems with relatively less shear remain in a sequence for the region $\delta\in(0.5,1.0)$.
After that, the tilted stellar objects with higher shearing motion can likely be observed near the parametric values $\delta=1$.
However, in GR, stellar objects with relativity less shearing
effects within the system are noticed. One can observe this kind of structure formation from the Figure \ref{shear1}.

\begin{figure} \centering
\epsfig{file=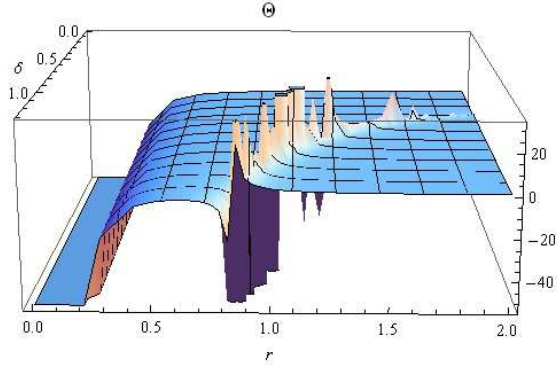,width=0.8\linewidth} \caption{Variation of the expansion scalar with respect to $\delta$ and
$r$.}\label{ex1}
\end{figure}

On the other hand, we observed that for the parametric choices of $\delta\in(0.4,1.0)$, the system having unit radius undergoes imploding and exploding phases due to the negative and positive values of $\Theta$. However, the same object enters into the stable window on taking $\delta=0$, i.e., GR case. This also indicates the occurrences of irregular distribution of matter due to inclusion of Palatini $f(R)$ factor $\delta$
as seen by Figure \ref{ex1}. All our findings and observations would reduce to GR
\cite{herrera2011tilted} under the limit $f(R)=R$.

\vspace{0.3cm}

\section*{Acknowledgments}

This work was supported financially by National Research Project for
Universities (NRPU), Higher Education Commission Pakistan under the
research Project No. 8754/Punjab/NRPU/R\&D/HEC/2017.

\vspace{0.3cm}

\renewcommand{\theequation}{A\arabic{equation}}
\setcounter{equation}{0}
\section*{Appendix}

The expressions $\mathcal{K}_0$ and $\mathcal{K}_1$ arising in
Eqs.\eqref{50} and \eqref{51} are given below
\begin{align}\nonumber
\mathcal{K}_0&=\frac{\mathcal{T}_{01}}{C^2}\left(\frac{2B'}{B}
+\frac{2f'_R}{f_R}+\frac{C'}{C}\right)-\dot{\mathcal{T}}_{00}
-\mathcal{T}_{00}\left(\frac{\dot{C}}{C}+\frac{3\dot{f_R}}{2f_R}
+\frac{2\dot{B}}{B}\right)-\frac{\mathcal{T}_{11}}{C^2}\\\label{A1}
&\times\left(\frac{\dot{C}}{C}+\frac{\dot{f_R}}{2f_R}\right)
+\left(\frac{\mathcal{T}_{10}}{C^2}\right)'-\frac{2\mathcal{T}_{22}}{B^2}\left(\frac{\dot{B}}{B}
+\frac{\dot{f_R}}{2f_R}\right),\\\nonumber
\mathcal{K}_1&=-\dot{\mathcal{T}}_{10}
+\mathcal{T}_{00}\frac{f_R'}{2f_R}+
\left(2\frac{B'}{B}+\frac{3f_R'}{2f_R}\right)\frac{\mathcal{T}_{11}}{C^2}-\mathcal{T}_{10}
\left(\frac{\dot{C}}{C}+\frac{2\dot{f}_R}{f_R}+\frac{2\dot{B}}{B}\right)\\\label{A2}
&+\left(\frac{\mathcal{T}_{11}}{C^2}\right)'-\frac{2}{B^2}\left(\frac{B'}{B}+\frac{f'_R}{2f_R}\right)\mathcal{T}_{22}.
\end{align}


\vspace{0.5cm}

\begin{thebibliography}{10}

\bibitem{pietrobon2006integrated}
D.~Pietrobon, A.~Balbi, and D.~Marinucci {\em Phys. Rev. D}, vol.~74,
  p.~043524, 2006.

\bibitem{giannantonio2006high}
T.~Giannantonio {\em et~al.} {\em Phys. Rev. D}, vol.~74, p.~063520, 2006.

\bibitem{riess2007new}
A.~G. Riess {\em et~al.} {\em Astrophys. J.}, vol.~659, p.~98, 2007.

\bibitem{guth1981inflationary}
A.~H. Guth {\em Phys. Rev. D}, vol.~23, p.~347, 1981.

\bibitem{peebles1993principles}
P.~J.~E. Peebles, {\em Principles of physical cosmology}.
\newblock Princeton university press, 1993.

\bibitem{komatsu2009five}
E.~Komatsu {\em et~al.} {\em Astrophys. J. Supp. Ser.}, vol.~180, p.~330, 2009.

\bibitem{nojiri2007introduction}
S.~Nojiri and S.~D. Odintsov {\em Int. J. Geom. Meth. Mod. Phys.}, vol.~4,
  p.~115, 2007.

\bibitem{NOJIRI2006144}
S.~Nojiri and S.~D. Odintsov {\em Phys. Lett. B}, vol.~639, p.~144, 2006.

\bibitem{Bamba2012}
K.~Bamba, S.~Capozziello, S.~Nojiri, and S.~D. Odintsov {\em Astrophys. Space
  Sci.}, vol.~342, p.~155, 2012.

\bibitem{capozziello2011extended}
S.~Capozziello and M.~De~Laurentis {\em Phys. Rep.}, vol.~509, p.~167, 2011.

\bibitem{bamba2013modified}
K.~Bamba, S.~Nojiri, and S.~D. Odintsov {\em arXiv preprint arXiv:1302.4831},
  2013.

\bibitem{yousaf2016causes}
Z.~Yousaf, K.~Bamba, and M.~Z. Bhatti {\em Phys. Rev. D}, vol.~93, p.~124048,
  2016.

\bibitem{yousaf2016influence}
Z.~Yousaf, K.~Bamba, and M.~Z. Bhatti {\em Phys. Rev. D}, vol.~93, p.~064059,
  2016.

\bibitem{nojiri2008dark}
S.~Nojiri and S.~D. Odintsov {\em arXiv preprint arXiv:0807.0685}, 2008.

\bibitem{nojiri2002friedmann}
S.~Nojiri, S.~D. Odintsov, and S.~Ogushi {\em Int. J. Mod. Phys. A}, vol.~17,
  pp.~4809--4870, 2002.

\bibitem{yousaf2019role}
Z.~Yousaf {\em Eur. Phys. J. Plus}, vol.~134, p.~245, 2019.

\bibitem{meng2004cosmological}
X.~H. Meng and P.~Wang {\em Class. Quantum Grav.}, vol.~21, p.~951, 2004.

\bibitem{allemandi2004accelerated}
G.~Allemandi, A.~Borowiec, and M.~Francaviglia {\em Phys. Rev. D}, vol.~70,
  p.~043524, 2004.

\bibitem{allemandi2007constraining}
G.~Allemandi and M.~L. Ruggiero {\em Gen. Relativ. Gravit.}, vol.~39, p.~1381,
  2007.

\bibitem{sharif2015stability}
M.~Sharif and Z.~Yousaf {\em Eur. Phys. J. C}, vol.~75, p.~58, 2015.

\bibitem{olmo2011palatini}
G.~J. Olmo {\em Int. J. Mod. Phys. D}, vol.~20, pp.~413--462, 2011.

\bibitem{sotiriou2006f}
T.~P. Sotiriou {\em Class. Quantum Grav.}, vol.~23, p.~5117, 2006.

\bibitem{PhysRevD.72.083505}
G.~J. Olmo {\em Phys. Rev. D}, vol.~72, p.~083505.

\bibitem{doi:10.1142/S0218271811018925}
G.~J. Olmo {\em Int. J. Mod. Phys. D}, vol.~20, p.~413, 2011.

\bibitem{PhysRevD.86.044014}
G.~J. Olmo and D.~Rubiera-Garcia {\em Phys. Rev. D}, vol.~86, p.~044014, 2012.

\bibitem{PhysRevD.86.104039}
G.~J. Olmo, H.~Sanchis-Alepuz, and S.~Tripathi {\em Phys. Rev. D}, vol.~86,
  p.~104039, 2012.

\bibitem{PhysRevD.86.127504}
S.~Capozziello, T.~Harko, T.~S. Koivisto, F.~S.~N. Lobo, and G.~J. Olmo {\em
  Phys. Rev. D}, vol.~86, p.~127504, 2012.

\bibitem{olmo2020stellar}
G.~J. Olmo, D.~Rubiera-Garcia, and A.~Wojnar {\em Phys. Rep.}, vol.~876, p.~1,
  2020.

\bibitem{olmo2020junction}
G.~J. Olmo and D.~Rubiera-Garcia {\em Class. Quantum Grav.}, 2020.

\bibitem{Ilyas2017}
M.~Ilyas, Z.~Yousaf, M.~Z. Bhatti, and B.~Masud {\em Astrophys. Space Sci.},
  vol.~362, p.~237, 2017.

\bibitem{moraes2017analytical}
P.~H. R.~S. Moraes, R.~A.~C. Correa, and R.~V. Lobato {\em J. Cosmol.
  Astropart. Phys.}, vol.~2017, p.~029, 2017.

\bibitem{bhatti2020dynamical}
M.~Z. Bhatti, Z.~Yousaf, and A.~Khadim {\em Phys. Review D}, vol.~101,
  p.~104029, 2020.

\bibitem{sahoo2017wormholes}
P.~K. Sahoo, P.~H. R.~S. Moraes, and P.~Sahoo {\em arXiv preprint
  arXiv:1709.07774}, 2017.

\bibitem{bambi2016wormholes}
C.~Bambi, A.~Cardenas-Avendano, G.~J. Olmo, and D.~Rubiera-Garcia {\em Phys.
  Rev. D}, vol.~93, p.~064016, 2016.

\bibitem{yousaf2020construction}
Z.~Yousaf {\em Phys. Dark Universe}, vol.~28, p.~100509, 2020.

\bibitem{yousaf2020gravastars}
Z.~Yousaf, M.~Z. Bhatti, and H.~Asad {\em Phys. Dark Universe}, vol.~28,
  p.~100527, 2020.

\bibitem{yadav2020existence}
A.~K. Yadav, L.~K. Sharma, B.~K. Singh, and P.~K. Sahoo {\em New Astr.},
  vol.~78, p.~101382, 2020.

\bibitem{malik2020study}
A.~Malik and M.~F. Shamir {\em New Astr.}, vol.~80, p.~101422, 2020.

\bibitem{yousaf2017stellar}
Z.~Yousaf {\em Eur. Phys. J. Plus}, vol.~132, p.~276, 2017.

\bibitem{hawking1979general}
S.~W. Hawking and W.~Israel.
\newblock CUP Archive, 1979.

\bibitem{herrera1997role}
L.~Herrera, A.~Di~Prisco, J.~Hern{\'a}ndez-Pastora, and N.~Santos {\em Phys.
  Lett. A}, vol.~237, p.~113, 1998.

\bibitem{herrera2004spherically}
L.~Herrera, A.~Di~Prisco, J.~Martin, J.~Ospino, N.~O. Santos, and O.~Troconis
  {\em Phys. Rev. D}, vol.~69, p.~084026, 2004.

\bibitem{herrera2011role}
L.~Herrera, A.~Di~Prisco, and J.~Ib\'{a}\~{n}ez {\em Phys. Rev. D}, vol.~84,
  p.~107501, 2011.

\bibitem{yousaf2017evolution}
Z.~Yousaf, M.~Z. Bhatti, and A.~Rafaqat {\em Can. J. Phys.}, vol.~95,
  pp.~1246--1252, 2017.

\bibitem{di2011expansion}
A.~Di~Prisco, L.~Herrera, J.~Ospino, N.~O. Santos, and V.~M. Vi{\~n}a-Cervantes
  {\em Int. J. Mod. Phys. D}, vol.~20, p.~2351, 2011.

\bibitem{bhatti2017evolution}
M.~Z. Bhatti, Z.~Yousaf, and M.~Ilyas {\em Eur. Phys. J. C}, vol.~77, p.~690,
  2017.

\bibitem{yousaf2018structure}
Z.~Yousaf {\em Astrophys. Space Sci.}, vol.~363, p.~226, 2018.

\bibitem{RYousaf2019}
Z.~Yousaf, M.~Z. Bhatti, and R.~Saleem {\em Eur. Phys. J. Plus}, vol.~134,
  p.~142, 2019.

\bibitem{herrera2011tilted}
L.~Herrera, A.~Di~Prisco, and J.~Ib{\'a}{\~n}ez {\em Phys. Rev. D}, vol.~84,
  p.~064036, 2011.

\bibitem{herrera2017gibbs}
L.~Herrera {\em Entropy}, vol.~19, p.~110, 2017.

\bibitem{herrera2020landauer}
L.~Herrera {\em Entropy}, vol.~22, p.~340, 2020.

\bibitem{PhysRevD.95.024024}
Z.~Yousaf, K.~Bamba, and M.~Z. Bhatti {\em Phys. Rev. D}, vol.~95, p.~024024,
  2017.

\bibitem{yousaf2019tilted}
Z.~Yousaf, M.~Z. Bhatti, and S.~Yaseen {\em Eur. Phys. J. Plus}, vol.~134,
  p.~487, 2019.

\bibitem{yousaf2019non}
Z.~Yousaf, M.~Z. Bhatti, and M.~F. Malik {\em Eur. Phys. J. Plus}, vol.~134,
  p.~470, 2019.

\bibitem{sussman2017lemaitre}
R.~A. Sussman and L.~G. Jaime {\em Class. Quantum Grav.}, vol.~34, p.~245004,
  2017.

\bibitem{doi:10.1142/S0217732319503334}
Z.~Yousaf {\em Mod. Phys. Lett. A}, vol.~34, p.~1950333, 2019.

\bibitem{starobinsky1980new}
A.~A. Starobinsky {\em Phys. Lett. B}, vol.~91, p.~99, 1980.

\bibitem{PhysRevD.75.127502}
I.~Sawicki and W.~Hu {\em Phys. Rev. D}, vol.~75, p.~127502, 2007.

\bibitem{ellis2012relativistic}
G.~F.~R. Ellis, R.~Maartens, and M.~A.~H. MacCallum, {\em Relativistic
  cosmology}.
\newblock Cambridge University Press, 2012.

\bibitem{herrera2011physical}
L.~Herrera {\em Int. J. Mod. Phys. D}, vol.~20, p.~1689, 2011.

\bibitem{Herrera_1997}
L.~Herrera, A.~D. Prisco, J.~L. Hern{\'{a}}ndez-Pastora, J.~Mart{\'{\i}}n, and
  J.~Mart{\'{\i}}nez {\em Class. Quantum Grav.}, vol.~14, p.~2239, 1997.

\bibitem{HERRERA2002157}
L.~Herrera {\em Phys. Lett. A}, vol.~300, p.~157, 2002.

\bibitem{PhysRevD.70.084004}
L.~Herrera and N.~O. Santos {\em Phys. Rev. D}, vol.~70, p.~084004, 2004.

\bibitem{yousaf2018some}
Z.~Yousaf and M.~Z. Bhatti {\em Int. J. Geom. Meth. Mod. Phys.}, vol.~15,
  p.~1850160, 2018.

\bibitem{herrera2009expansion}
L.~Herrera, G.~Le~Denmat, and N.~O. Santos {\em Phys. Rev. D}, vol.~79,
  p.~087505, 2009.

\end{thebibliography}
\end{document}